\documentclass[aps,twocolumn,showpacs,preprintnumbers,nofootinbib,tightenlines]{revtex4}
\usepackage{bm}% bold math
\usepackage{psfig}
\usepackage{times}
\newcommand{\be}{\beta}

\def\be{\begin{equation}}
\def\ee{\end{equation}}
\begin{document}
\title{Electrically charged compact stars and formation of charged black holes}

\author{Subharthi Ray}
\email{sray@if.uff.br}
\author{Aquino L. Esp\'{\i}ndola}
\email{aquino@if.uff.br}
\author{Manuel Malheiro}
\email{mane@if.uff.br}
\affiliation{Instituto de Fisica, Universidade Federal Fluminense,
Niteroi 24210-340, RJ, Brazil}
\author{Jos\'e P. S. Lemos}
\email{lemos@physics.columbia.edu}
\affiliation{Department of Physics, Columbia University, 
New York, NY 10027, USA, 
\&\\
Centro Multidisciplinar de Astrof\'{\i}sica - CENTRA,
Departamento de F\'{\i}sica, Instituto Superior T\'ecnico,
Av. Rovisco Pais 1, 1049-001 Lisboa, Portugal}
\author{Vilson T. Zanchin}
\email{zanchin@ccne.ufsm.br}
\affiliation{Universidade Federal Santa Maria, Departamento de Fisica,
BR-97105-900, Santa Maria, RS, Brazil}

\date{today}

\begin{abstract}
{We study the effect of electric charge in compact stars assuming that
the charge distribution is proportional to the mass density. The
pressure and the density of the matter inside the stars are large, and
the gravitational field is intense.  This indicates that electric
charge and a strong electric field can also be present. The
relativistic hydrostatic equilibrium equation, i.e., the
Tolman-Oppenheimer-Volkoff equation, is modified in order to include
electric charge.  We perform a detailed numerical study of the effect
of electric charge using a polytropic equation of state. We conclude that
in order to see any appreciable effect on the phenomenology of the
compact stars, the electric fields have to be huge ($\sim 10^{21}$
V/m), which implies that the total charge is $Q\sim 10^{20}\,$Coulomb.
From the local effect of the forces experienced on a single charged
particle, it is expected that each individual charged particle is
quickly ejected from the star. This in turn produces a huge force
imbalance, and the gravitational force overwhelms the repulsive
Coulomb and fluid pressure forces. The star can then collapse to form a
charged black hole before all the charge leaves the system.}
\end{abstract}


\maketitle

\section{Introduction}

The study of the effect of electric charge and electric field in a
gravitationally bound system has been done previously by some authors.
Rosseland, in 1924 \cite{ross24} (see also Eddington \cite{ed26}),
studied the possibility that a self gravitating star on Eddington's
theory, where the star is modeled by a ball of hot ionized gas, did
contain a net charge.  In such a system the electrons (lighter
particles) tend to rise to the top because of the difference in the
partial pressure of electrons compared to that of ions (heavier
particles).  The motion of electrons to the top and further escape
from the star is stopped by the electric field created by the charge
separation. The equilibrium is attained after some amount of electrons
escape leaving behind an electrified star whose net positive charge is
of about 100 Coulomb per solar mass, and building an interstellar gas
with a net negative charge. As shown by Bally and Harrison
\cite{bally78}, this result applies to any bound system whose size is
smaller than the Debye length of the surrounding media.  In fact, one
should expect a star like the sun to hold some amount of net charge
due to the much more frequent escape of electrons than that of
protons.  Moreover, one should also expect that the escape would stop
when the electrostatic energy of an electron $e\Phi$ is of the order
of its thermal energy $kT$. This gives for a ball of hot matter with
the sun radius, a net charge $Q \sim 6.7\times 10^{-6}T$ (in
Coulomb). Hence, the escape effect cannot lead to a large net electric
charge, and the conclusion is that a star formed by an initially
neutral gas cannot acquire a net electric charge larger than about
$100$C per solar mass.

In the case of cold stars, the gravitational pull is balanced by the
degeneracy pressure of the particles.  One can ask whether a net
charge can have any effect on the structure of the system. For a star
of mass $M$ and charge $Q$, the electrostatic energy of a particle at
radius $r$, $e Q/r$, is balanced by its gravitational energy, $mM/r$,
where $e$ and $m$ are the charge and mass of the particle, say a
proton.  For a spherical ball this gives a charge of approximately
$100$C per solar masses (see Glendenning\cite{glen00}). This is
clearly so for Newtonian stars. For compact stars, the high density
and relativistic effects must be taken into account in order to reproduce
with precision, the phenomenologies such as the mass and the radius.
These effects also reflect, in principle, in
the allowed net charge of a compact star, the star can take some more
charge to be in equilibrium (see e.g. Bekenstein \cite{bek71}).

That the observed stars, like the Sun or a neutron star, in
equilibrium cannot support a great amount of charge comes from the
fact that the particles that compose a star have a huge charge to mass
ratio, as is the case for a proton or an electron. For highly
relativistic stars, full general relativity is necessary, and the
situation might be different.  One can have highly compact stars,
whose radius is on the verge of forming an event horizon, such that
the huge gravitational pull can be balanced by huge amounts of net
charge. This type of configurations were raised by Bekenstein
\cite{bek71} and further developed in the studies by Zhang et
al. \cite{zhang82}, de Felice and Yu \cite{fel95}, de Felice et
al. \cite{fel99}, Yu and Liu \cite{yu00}, Anninos and Rothman
\cite{ani01} and others. For instance, Zhang et al. \cite{zhang82}
found that the structure of a neutron star, for a degenerate
relativistic fermi gas, is significantly affected by the electric
charge just when the charge density is close to the mass density (in
geometric units).  In the investigations by de Felice et al., and by
Anninos and Rothman, they assumed that the charge distribution
followed particular functions of the radial coordinate.

In passing, we can mention that if the charge to mass ratio of the
particles that make the star is low, say one or of order one, then a
star can contain a huge amount of charge. For instance, one could think
of dust particles containing a large quantity of neutral particles,
say $10^{18}$ neutrons, and one proton. This would give a charge to
mass ratio of the order one. The dust particles could then cluster
around each other and form a star.  Theoretically these stars have
been considered.  The Newtonian theory of gravitation and
electrostatics admit equilibrium configurations of charged fluids
where the charge density $\rho_{\rm ch}$ can be as large as the mass
density $\rho$, in appropriate units. For instance, a static
continuous distribution of charged dust matter with zero pressure will
be in equilibrium everywhere if $\rho_{\rm ch}= \pm\rho$ (in geometric
units). This gives for a ball of dust with the mass of the sun a net
charge of about $1.7 \times 10^{20}$ Coulomb.  The general
relativistic analog for charged dust stars was discovered by Majumdar
\cite{maj47} and by Papapetrou \cite{papa47}, and further discussed by
Bonnor \cite{bon75} and several other authors (see \cite{ivanov02c}
for a review). We will not pursue this line here.

In this paper we shall take the ideas initiated in \cite{bek71}
and continue in \cite{zhang82,fel95,fel99,yu00,ani01}, and
focus attention on the effect of charge in cold
compact stars, made of neutrons, protons and electrons, say.
We assume that the charge density goes with mass
density and write $\rho_{\rm ch}=\alpha \rho$, $\alpha$ being a
positive (negative) constant for a positive (negative) charge density.
We further assume that the net charge in the system is in the form of
trapped charged particles carrying positive electric charge.  Later we
will show that the negative charge will also give the same results,
thus the sign of the charge has no effects.  From the Einstein-Maxwell
equations, we write a modified Tolman-Oppenheimer-Volkoff (TOV)
equation (see \cite{bek71}) where the energy density which appears
from the electrostatic field will add up to the total energy density
of the system, which in turn will help in the gaining of the total
mass of the system. The fundamental difference to the standard
relativistic TOV equation from the uncharged case is the presence of
the Maxwell stress tensor, which in the modified TOV equation
contributes to the energy density, and to the pressure, besides the
presence of a potential gradient term from the coulombian term.
We solve the modified TOV equation
for  polytropic equations of  state assuming  that the  charge density
goes with the  matter density and discuss the  results. We will  show
that a previous estimate  by  Bekenstein is correct \cite{bek71}:
in order to see any appreciable effect on the phenomenology of
the neutron  stars, the  charge and the  electrical fields have  to be
huge, the large electric field being able to start producing
pair creation of particles.

Furthermore, we find solutions with huge amounts of charge.  How this
extra charge is formed in the star is not the concern here.  A
mechanism to generate charge asymmetry for charged black holes has
been suggested recently \cite{herman03} and the same may be applied
for compact stars too.  We are not claiming that compact stars always
have so large a charge and strong electric field.  Indeed, one has to
raise the question what happens to a single charged particle inside
the system. It will face a huge electrostatic force and will soon
escape the system making the star unstable.  During that process, the
gravitational force, which was previously balanced to some extent by
the repulsive Coulombian force, will overpower the inside pressure and
the system will collapse to form a black hole. However, not all the
charge has time to escape from the system and hence the charge will
remain trapped inside the event horizon to form a charged black hole.
This scenario could be applicable for newly born charged compact stars
and their rapid transformation to charged black holes.

This paper is organized in the following sections.  In section
\ref{sec:rel-form}, we give the general relativistic formulation for
the inclusion of charge in the TOV equation and present the numerical
procedure to solve them for a given radial charge distribution inside
the star.  Section \ref{sec:poly} presents the application of these
equations to polytropic stars wherein different aspects of the effect
of charge are described, such as mass-radius-charge diagrams, the
inside electric fields and the metric.  In section \ref{sec:ns2bh}, we
discuss the stability of these stars and the possibility of formation
of charged black holes. Finally we draw our conclusions in section
\ref{sec:conc}.

\section{General Relativistic Formulation}\label{sec:rel-form}

We take the metric for our static spherical star as
\begin{equation}
ds^2=e^\nu c^2 dt^2 - e^\lambda dr^2 - r^2(d\theta^2 + \sin^2\theta d\phi^2).
\label{eq:1}
\end{equation}

The stress tensor $T^\mu_\nu$ will include the terms from the Maxwell's
equation and the
complete form of the Einstein-Maxwell stress tensor will be :
\begin{equation}
T^\mu_\nu = (P + \epsilon)u^\mu u_\nu - P \delta^\mu_\nu + \frac{1}{4\pi}
\left(F^{\mu\alpha}F_{\alpha\nu} - \frac{1}{4}\delta^\mu_\nu F_{\alpha\beta}
F^{\alpha\beta}\right)
\label{eq:tmunu}
\end{equation}
where $P$ is the pressure,  $\epsilon$ is  the energy  density (=$\rho
c^2$) and  $u^\mu$ is  4-velocity vector. For the  time component,
one  easily  sees  that  $u^t  = e^{-\nu/2}/c$  and  hence  $u^tu_t=1.$
Consequently, the other components  (radial and spherical) of the four
vector are absent.

Now, the electromagnetic field is taken from the Maxwell's field equations
and hence they will follow the relation
\begin{equation}
\left[ \sqrt{-g} F^{\mu\nu}\right]_{,\nu}~=~4 \pi j^\mu \sqrt{-g}
\label{eq:fmunu}
\end{equation}
where $j^\mu$ is the four-current density. Since the present choice of
the  electromagnetic  field  is  only  due to  charge,  we  have  only
$F^{01}=-F^{10}$, and the other terms  are absent.  In general, we can
derive  the   electromagnetic  field  tensor   $F^{\mu\nu}$  from  the
four-potential  $A_\mu$.  So,  for  non vanishing  field  tensor,  the
surviving  potential  is  $A_0=\phi$.  We  also  considered  that  the
potential has a spherical symmetry, i.e., $\phi=\phi(r)$.

The non vanishing term in Eq. (\ref{eq:fmunu}) is when $\nu=r$. This gives the
electric field as
\begin{eqnarray}
\nonumber
^t_t~\sim~ ^r_r~:~\frac{1}{4\pi}
\left(F^{\mu\alpha}F_{\alpha\nu} - \frac{1}{4}\delta^\mu_\nu F_{\alpha\beta}
F^{\alpha\beta}\right)=\frac{-{\cal U}^2}{8\pi}
\end{eqnarray}
where,
\begin{equation}
{\cal U}(r)=\frac{1}{r^2}\int^r_0 4 \pi r^2 j^0 e^{(\nu + \lambda)/2} dr
\label{eq:ucal1}
\end{equation}
is the electric field. Let us consider a distribution of charge
in the matter as some charge density annoted as $\rho_{\rm ch}$.
We modify the above expression (Eq. (\ref{eq:ucal1}))  and
rewrite it as
\begin{equation}
{\cal U}(r)=\frac{1}{r^2}\int^r_0 4 \pi r^2 \rho_{\rm ch} e^{\lambda/2} dr.
\label{eq:ucal2}
\end{equation}
This leads to defining the total charge of the system as
\begin{equation}
Q=\int^R_0 4 \pi r^2 \rho_{\rm ch} e^{\lambda/2} dr
\label{eq:charge}
\end{equation}
where $R$ is the radius of the star.

With the  metric (\ref{eq:1}) one  can easily get the  Einstein's field
equations from the relation
$$ R^\mu_\nu - \frac{1}{2}R \delta^\mu_\nu = -\frac{8\pi G}{c^4}T^\mu_\nu $$ as
\begin{equation}
^t_t~:~ \frac{e^{-\lambda}}{r^2}\left(r\frac{d\lambda}{dr}-1\right) +
  \frac{1}{r^2} = \frac{8\pi G}
{c^4}\left(\epsilon + \frac{{\cal U}^2}{8\pi}\right)~~~~~~~~~~~~~
\end{equation}
\begin{equation}
^r_r~:~ \frac{e^{-\lambda}}{r^2}\left(r\frac{d\nu}{dr}+1\right) -
 \frac{1}{r^2}\\
= \frac{8\pi G}
{c^4}\left(P - \frac{{\cal U}^2}{8\pi}\right).
\end{equation}

The first of the Einstein's  equations is used to determine the metric
$e^\lambda$. The mass of the star is now due to the total contribution
of  the energy  density of the  matter and  the electric  energy ($\frac{{\cal
U}^2}{8\pi}$) density. The mass takes the new form as
\begin{equation}
M_{\rm tot}(r)=\int^r_04\pi r^2\left(\frac{\epsilon}{c^2}+
\frac{{\cal U}^2}{8\pi c^2}\right)dr
\label{eq:mnew}
\end{equation}
and the metric coefficient is given by
\begin{equation}
e^{-\lambda}=1-\frac{2GM_{\rm tot}(r)}{c^2r}.
\label{eq:lamnew}
\end{equation}
However, this mass is the one measured in the star's frame.
For an observer at infinity, the mass \cite{bek71,mac01} is given by
\begin{eqnarray}
\nonumber
M_\infty &=&\int^\infty_04\pi r^2\left(\frac{\epsilon}{c^2}+
\frac{{\cal U}^2}{8\pi c^2}\right)dr\\
\nonumber
&=&\int^R_04\pi r^2\left(\frac{\epsilon}{c^2}+
\frac{{\cal U}^2}{8\pi c^2}\right)dr +\\
\nonumber
&& \int^\infty_R4\pi r^2\left(\frac{\epsilon}{c^2}+
\frac{{\cal U}^2}{8\pi c^2}\right)dr\\
&=&M_{\rm tot}(R) + \frac{Q(R)^2}{2R}
\label{eq:minfty}
\end{eqnarray}
where R is the radius of the star.
In our plots and rest of the text, we will refer to the mass $M$ as the
$M_{\rm tot}$.

From the conservation of stress tensor
(${{T_\nu}^\mu}_{;\mu}=0$)  one  gets   the  form   of the
hydrostatic equation. To this end, we obtain the modified TOV as
\begin{eqnarray}
\nonumber
\frac{dP}{dr}=-\frac{G\left[M_{\rm tot}(r)+4\pi r^3\left(\frac{P}{c^2}-
\frac{{\cal U}^2}{8\pi c^2}
\right)\right](\epsilon+P)}{c^2r^2\left(1-\frac{2GM_{\rm tot}}{c^2r}\right)}\\
~~~~~~~~~~~~~~~~~~~~~~~~~~~~
+\rho_{\rm ch}\,{\cal U}\,e^{\frac{\lambda}{2}}.
\label{eq:dpdr}
\end{eqnarray}
The first term on the r.h.s. comes form the gravitational force with
an effective pressure and density which we will discuss later, and the
second term from the Coulomb force that depends on the matter by the
metric coefficient.

Numerical solutions in fortran require the integral forms of Eqs.
(\ref{eq:ucal2}),
(\ref{eq:mnew}) \& (\ref{eq:lamnew}) to be expressed in their differential
forms. We can write the corresponding differential forms as
\begin{equation}
d{\cal U}=-\frac{2{\cal U}dr}{r}+4\pi\rho_{\rm ch}e^{\lambda/2}dr,
\label{eq:du}
\end{equation}
\begin{equation}
dM_{\rm tot}=4\pi r^2\left(\frac{\epsilon}{c^2}+\frac{{\cal U}^2}{8\pi
  c^2}\right)dr
\label{eq:dm}
\end{equation}
and
\begin{equation}
d\lambda=\left[\frac{8\pi G}{c^2}re^\lambda\left(\frac{\epsilon}{c^2}
+\frac{{\cal U}^2}{8\pi c^2}\right)-
\left(\frac{e^\lambda-1}{r}\right)\right]dr.
\label{eq:dl}
\end{equation}
So, the final four equations needed to be solved are Eqs. (\ref{eq:dpdr}),
(\ref{eq:du}),
(\ref{eq:dm}) \& (\ref{eq:dl}).
The boundary conditions for the solution are, at the centre where $r=0,$
${\cal U}(r) = 0,$ $e^{\lambda(r)}=1,$ $P(r)=P_c$ and $\rho(r)=\rho_c,$
and at the surface
where $r=R,$ $P(r)=0.$ The inputs in the equations are
the pressure $P$, the energy density $\epsilon$ and the charge density
$\rho_{\rm ch}$. The metric coefficient $\lambda$ and the electric field ${\cal U}$
are interdependent. This gives us a set of four coupled differential
equations which we solve simultaneously
to get our results. We note here that the form of the equations does not
change with the
sign of the charge because the electric field appears in the mass term
(\ref{eq:dm}) and
pressure gradient term (\ref{eq:dpdr}) in squares and in the Coulomb part,
the product
$\rho_{\rm ch}{\cal U}$ is also invariant.

\section{Charged stars in a Polytropic equation of state}\label{sec:poly}

\subsection{The Mass-Radius relation and other features}

We  now examine  the effects  of charge  in a  polytropic  equation of
state (EOS),  which is  a more  general approach  than  considering any
model dependent  EOS.  We make the  charge go with  the mass density
($\epsilon$) as
\begin{equation}
\rho_{\rm ch}=f\times\epsilon
\end{equation}
where $\epsilon=\rho c^2$ is in [MeV/fm$^3$]. With this assumption, the
charge fraction $f$ has
a dimension $\frac{1}{\rm [km]}[{\rm fm}^3/{\rm MeV}]^{1/2}$ and the charge density
$\rho_{\rm ch}$ is
in $[{\rm MeV/fm}^3]^{1/2}\frac{1}{\rm [km]}$. This kind of `mixed units' appear in our
dimensions of
$f$ and $\rho_{\rm ch}$ because one can see from Eqs. (\ref{eq:ucal2}) \&
(\ref{eq:dpdr}) that
the electric field is proportional to the square root of the pressure (in
units of MeV/fm$^3$) and the integration
over the radius $r$ is carried out in kilometers. In geometrical units, this
can be written as
\begin{equation}
\rho_{\rm ch}=\alpha \times\rho
\label{eq:no-dim}
\end{equation}
where charge is expressed in units of mass and charge density in units of
mass density.
This $\alpha$ is related to our charge fraction $f$ as
\begin{eqnarray}
\nonumber
\alpha&=&f\times\frac{0.224536}{\sqrt{G}}\\
&=&f\times 0.86924 \times 10^3.
\label{eq:no-dim-conv}
\end{eqnarray}
Our choice of charge distribution  is a reasonable assumption
in the sense that large mass can hold a large amount of charge.

The polytropic EOS is given by
\begin{equation}
P=\kappa \rho^{1+1/n}
\end{equation}
where  $n$ is the  polytropic index  and is  related to  the
exponent  $\Gamma$  as  $n=\frac{1}{\Gamma  -1}$. In  the
relativistic regime,   the  allowed   value   of  $\Gamma$   is
$\frac{4}{3}$   to $\frac{5}{3}$.   We   have    considered  the
case   of $\Gamma=\frac{5}{3}$   and   the  corresponding
value   of  $n$   is 1.5. Primarily,  our  units of  matter
density  and  pressure are  in MeV/fm$^3$. We choose a value of
$\kappa$ as 0.05 $ [fm]^{8/3}$. Thus we have an equation
of state which we analyze for different cases of charge fraction
$f$ and study the nature and behavior of the system. The choice
of the EOS is however not very important in so far as the nature
of the curves due to the effect of charge is concerned, and show
only corresponding shift in the maximum mass based on the type of
EOS used.

It should be noted that the nuclear forces are only affected by
electromagnetic forces (and so the EOS) when the number density
of the charged particles are of the order of the baryonic number
density, i.e., Z $\sim$ A. This has been verified previously in
many works on the nuclear structure. However, for the case of
charged stars, the forces are compared between the electric and
the gravitational force. For our case, the Z/A ratio is
$\sim~10^{-18}$, which essentially means that these extra charged
particles which produces huge electric field and affects the
structure of the star, produces negligible effect on the nuclear
matter and the EOS. The same argument holds for the chemical
potential, which are controlled by chemical equilibrium and
charge neutrality. The extra charged particles are just very few
when compared to the total number of baryons of the star. Any
effect brought in by these extra charged particles can be
approximated (and compared) to the case of chemical equilibrium
and zero charge. So, it is justified to use an EOS which is
calculated for neutral matter. We have indeed tested the effect
of this amount on charge on the EOS on some sophisticated models
of nuclear matter like the Non-linear Walecka model
\cite{taurines2001,glen92}, and verified our reasoning to be
right.

\begin{figure}[htbp]
\centerline{\psfig{figure=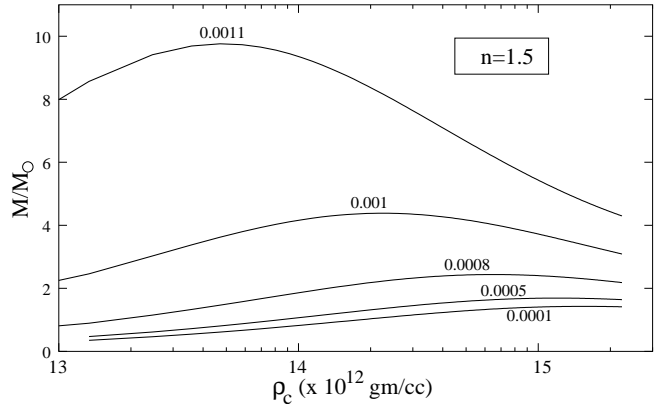,width=8.6cm}}\vskip .5cm
\caption{Central density against mass for different values of the factor $f$.}
\label{fig:poly-e-m}
\end{figure}

We plot the mass as a function of the central density $\rho_c$
in  Fig. (\ref{fig:poly-e-m})  for  different values  of the  charge
fraction $f$.  The stars  which are on  the higher density  regime and
have  lower mass, are  unstable because  $\frac{dM}{d\rho_c}<0$. Those
falling in the  lower density regime and have  increasing mass are all
allowed.  The effect of the  charge for $f=0.0001$ on the structure of
the star  is not  profound and is  comparable with that  of chargeless
star.  This value  of  $f$ is  however  `critical' in  the sense  that
further increase  in the  value shows effect  on the structure  of the
star.  With the  increase of  the charge  fraction form  $f=0.0001$ to
$f=0.0005$ the structure  changes by 20\% increase in  the value, from
$f=0.0005$ to $f=0.0008$, the increase  is 35\% and from $f=0.0008$ to
$f=0.001$, the change is almost  90\%, thus showing that the change in
the structure is non-linear with  the change in the charge fraction as
can be seen in Fig. (\ref{fig:poly-e-m}).

\begin{figure}[htbp]
\centerline{\psfig{figure=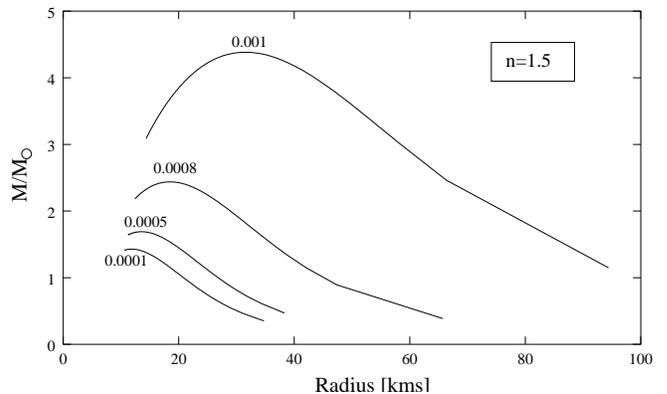,width=8.6cm}}\vskip .5cm
\caption{Mass-Radius relation for different values of the factor $f$.}
\label{fig:poly-m-r}
\end{figure}

In Fig. (\ref{fig:poly-m-r}) we plot the mass-radius relation.
Due to the effect of the repulsive force, the charged stars have
large radius and larger mass as we should expect. Even if the
radius is increasing with the mass, the M/R ratio is also
increasing, but much slower.  For the lower charge fractions,
this increase in the radius is very small, but a look at the
structure for the fraction $f=0.001$ reveals that for a mass of
4.3 M$_\odot$, the radius goes as high as 35 km. Though the
compactness of the stars are retained, they are now better to be
called as `compact charged stars' rather than `charged neutron
stars'. The charge fraction in the limiting case of maximally
allowed value goes up to $f=0.0011$, for which the maximum mass
stable star forms at a lower central density even smaller than
the nuclear matter density. This extreme case is not shown in
Fig. (\ref{fig:poly-m-r}) because the radius of the star and its
mass is very high (68 km and 9.7 M$_\odot$ respectively). These
effects suppress the curves of the lower charge fractions due to
scaling. For this star, the mass contribution from the electric
energy density is 10\% than that from the mass density.  It can
be checked by using relation (\ref{eq:no-dim-conv}) that this
charge fraction $f=0.0011$ corresponds to $\rho_{\rm
ch}=0.95616\times\rho$ in geometrical units.  In Table
\ref{tab:m-r}, we show the maximum mass for different charge
fractions and their corresponding radii, central densities and
net charge content.

\begin{table}[htp]
\caption{The maximum allowed stable stellar configuration for different charge fractions.}
\begin{center}
\begin{tabular}{|c|c|c|c|c|c|}
\hline
$f$ & M$_{\rm tot}$ & M$_\infty$ & R & $\epsilon_c$ & Q \\
 & (M$_\odot$) & (M$_\odot$) & (km)  & (MeV/fm$^3$) & ($\times 10^{20}$ Coulomb) \\
\hline
0.0001 &1.428 & 1.43  & 11.87 &1550.41 & 0.259 \\
0.0005 &1.69 & 1.765 & 13.55 &1202.6 & 1.517\\
0.0008 &2.438 & 2.728 & 18.47 & 652.87 & 3.434 \\
0.001  & 4.384 & 5.248 & 31.47 & 226.55 &  7.576\\
0.0011 & 9.76 & 12.15 & 68.7 &47.04 &18.314\\
\hline
\end{tabular}
\end{center}
\label{tab:m-r}
\end{table}

\begin{figure}[htbp]
\centerline{\psfig{figure=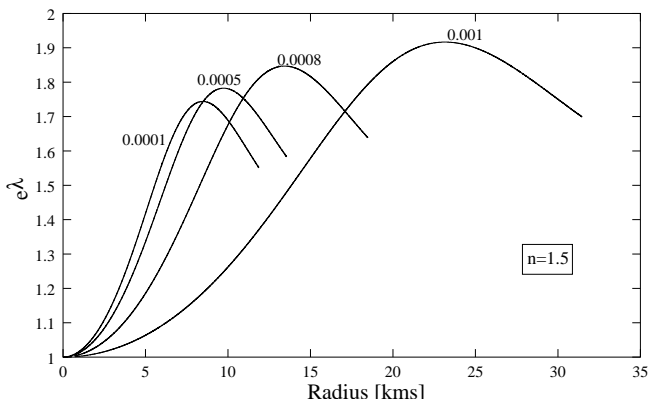,width=8.8cm}}\vskip .5cm
\caption{Variation of the metric e$^\lambda$ with the radius in the
maximum mass stars for different cases of charge distribution with varying
$f$.}
\label{fig:poly-l-r}
\end{figure}

In Fig. (\ref{fig:poly-l-r}), we plot the metric coefficient
e$^{\lambda}$ as a  function of radius for the  maximum mass stars for
each of the charge fractions. A quick comparison shows that the nature
of  e$^{\lambda}$  is  the  same  for all  the  stars  with  different
charges. There  is a  slight increase in  its value for  higher charge
fractions   thus    showing   the gain    in   the   compactness
$\frac{M}{R}$ of the star with charge. This can be verified from the
values of $e^\lambda$ at the surface for two cases of charge fractions
$f=0.0001$ and $f=0.001$, 
$\left(\frac{2M}{R}\right)_{0.0001}=1-\frac{1}{1.55} = 0.3548$, and 
$\left(\frac{2M}{R}\right)_{0.001}=1-\frac{1}{1.7} = 0.4118$. 
This re-confirms that the compactness of the star increases despite the
enormous  increase in  the  radius  with the  increase  of the  charge
fraction.

It is interesting to point out that we have tested our model of
charged stars for a very soft EOS which can go upto a very high
density, where we saw that the spiraling behaviour of the
mass-radius curve exists for high charges. This nature of the
curve are very much supportive to the fact that these charged
stars are stable to stellar oscillations.

\subsection{Charge and Electric field inside the star.}
\begin{figure}[htbp]\vskip .5cm
\centerline{\psfig{figure=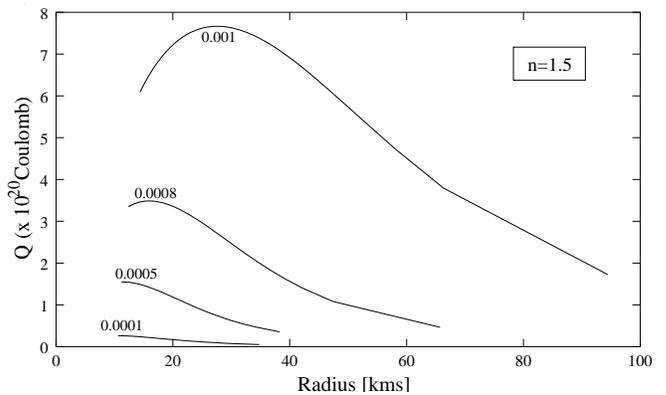,width=8.6cm}}\vskip .5cm
\caption{The variation of the charge with radius for different $f$.}
\label{fig:poly-q-r}
\end{figure}

In this  subsection we discuss the  effect of the charge  in the inner
profile of  the star.  In the $Q\times R$ diagram in
Fig. (\ref{fig:poly-q-r}),
we plot the
total  charge $Q$  of  the stars  at  the surface as function of their
radius. From Eq.  (\ref{eq:charge}), it is clear that  the charge will
increase with the increasing charge fraction. This increase is however
not  the only  one responsible  for the  high charges  if  we consider
that equation. There  are contributions  also from  the large
radius of the higher charge  configuration and the larger value of the
metric  coefficient e$^{\lambda}$.  So,  we see  that  the matter  and
charge are  inter-related to each other  and the effect  of one depends
intrinsically  on  the other  and  hence  the  relations are  strongly
coupled.

The charge  developed due to this range of charge fraction $(f=0.0001
\,{\rm to}\, f=0.0011)$ is  in the limiting
case where the  effect is noticed on the structure.  This scale of the
charge is easy to understand from the mass expression and the modified
TOV (Eqs. (\ref{eq:mnew}) and (\ref{eq:dpdr})) where in order to see
any change in the stellar configuration, we should have
\begin{eqnarray}
{\cal U} \simeq \sqrt{8\pi P} < \sqrt{8\pi\epsilon}.
\label{eq:uisless}
\end{eqnarray}
If  we  consider  that  $P\sim10 \,\, {\rm MeV/fm}^3$,
then  ${\cal  U}\simeq
10\left(\frac{\rm MeV}{{\rm fm}^3}\right)^{1/2}$
and we  show in  appendix that
with    proper     conversion,    this    gives     ${\cal    U}\simeq
10^{22}$V/m.  With  R$\simeq$10 km, the charge needs  to be at
the order of 10$^{20}$ Coulombs.

\begin{figure}[htbp]\vskip .5cm
\centerline{\psfig{figure=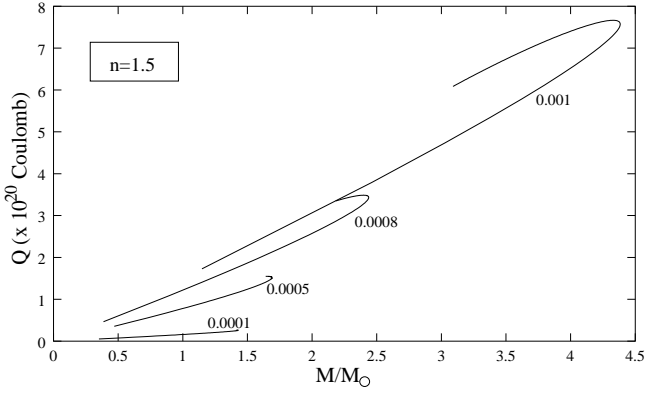,width=8.6cm}}\vskip .5cm
\caption{\label{fig:poly-q-m}The variation of the charge with mass for
  different $f$.}
\end{figure}

In the $Q\times M$ diagram in Fig. (\ref{fig:poly-q-m}), we plot
the mass of the stars against their surface charge. We have made
the charge density proportional to  the energy  density and so
it was expected  that the charge, which is  a volume integral of
the charge  density, will go in the same  way as the  mass, which
is  also a volume integral  over the mass density.  The slope of
the curves comes from the different charge fractions.The nature
of the curves  in fact reflect that charge varies with mass
(with the  turning back  of the curves  all falling  in the
`unstable zone' and is  not taken into consideration). If we
consider that the maximum charge allowed  is estimated  by the
condition (\ref{eq:uisless}) for $\frac{dP}{dr}$ to be negative
(Eq.~(\ref{eq:dpdr})),  we see that the  curve for the maximum
charge  in Fig. (\ref{fig:poly-q-m}) has a  slope of 1:1 (in a
charge scale of 10$^{20}$  Coulomb). This scale can be easily
understood if we write the charge as
\begin{eqnarray}
Q=\sqrt{G}M_\odot\frac{M}{M_\odot}\simeq10^{20}\frac{M}{M_\odot}
\, {\rm Coulomb}.
\end{eqnarray}
It is worth mentioning  that this charge $Q$  is the charge at
the surface of the  star where the  pressure and  also
$\frac{dP}{dr}$  are already very small  (ideally zero). So, at
the surface, the  Coulomb force is essentially balanced  by the
gravitational  force and the  relation of the charge and mass
distribution we  found is of the same order (i.e., $Q\simeq M~{\rm
or}~\rho_{\rm ch}\simeq\rho$) for the case  of   charged  dust
sphere  discussed  earlier   by  Papapetrou \cite{papa47} and
Bonnor \cite{bon75}, as we referred in the introduction.

\begin{figure}[htbp]\vskip .5cm
\centerline{\psfig{figure=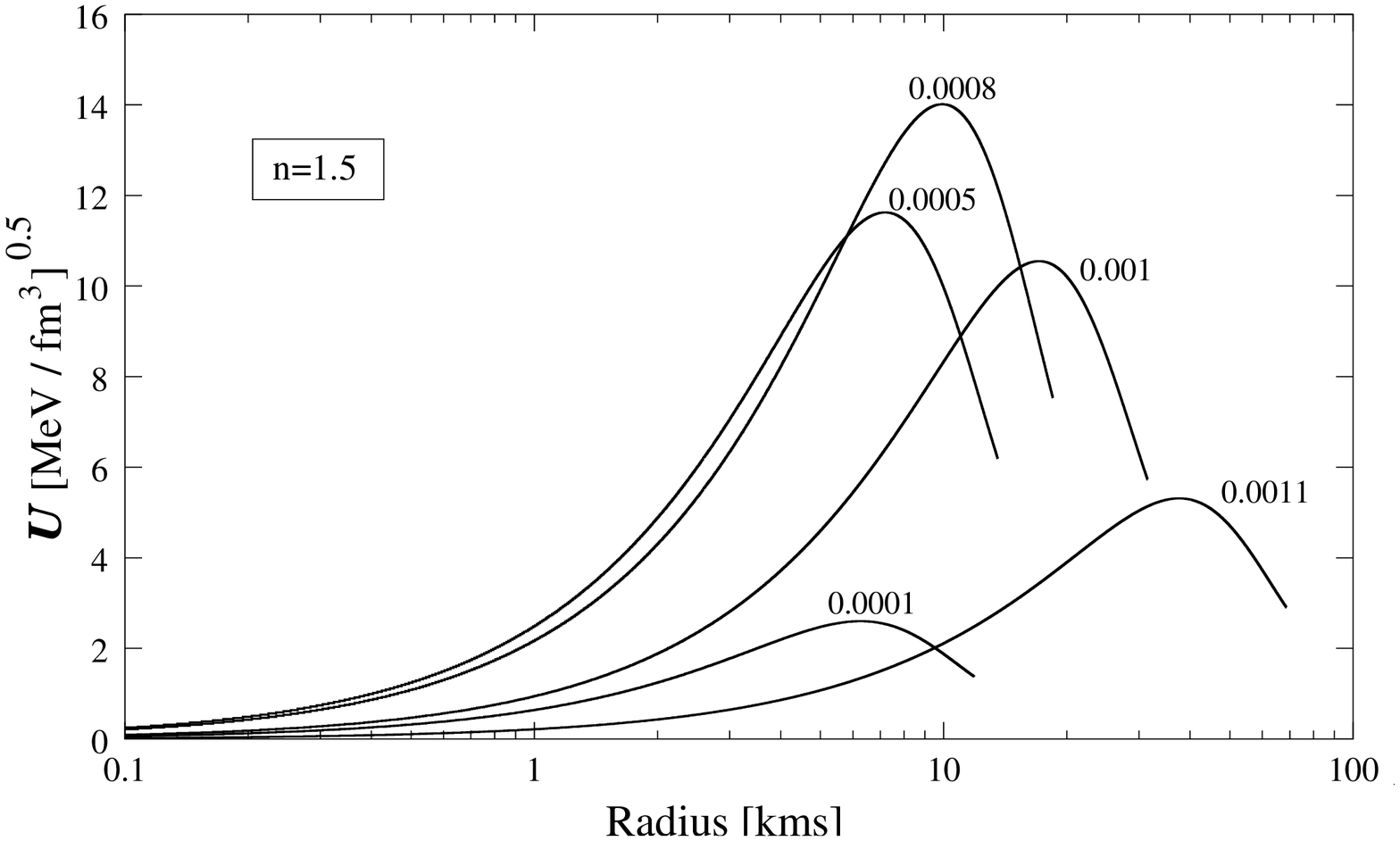,width=8.6cm}}\vskip .5cm
\caption{Variation of ${\cal U}$ with the radius in the
maximum mass stars for different cases of charge distribution with varying
$f$.}
\label{fig:poly-u-r}
\end{figure}

In  Fig. (\ref{fig:poly-u-r}), the  electric  field  ($\cal U$)  is
plotted as  a function of the  stellar radius for the maximum mass star
for different charge fraction $f$. The radius  is shown in
a  log-scale.  From the  expression  (Eq.  (\ref{eq:ucal2})) of  the
electric field, it is clear that  the profile of the field will depend
on the  charge fraction $f$, the metric  coefficient e$^{\lambda}$ and
also the  radius of  the star.  The value of  the field  increases for
charge fraction  up to $f=0.0008$ but  then falls down. The  decrease in the
field for the
higher charge  fraction is attributed  to the formation of  the stable
star at  a very small  density (Fig. (\ref{fig:poly-e-m}))  for that
particular charge fraction $f$.

\subsection{The modified EOS inside the stars}

\begin{figure}[htbp]\vskip .5cm
\centerline{\psfig{figure=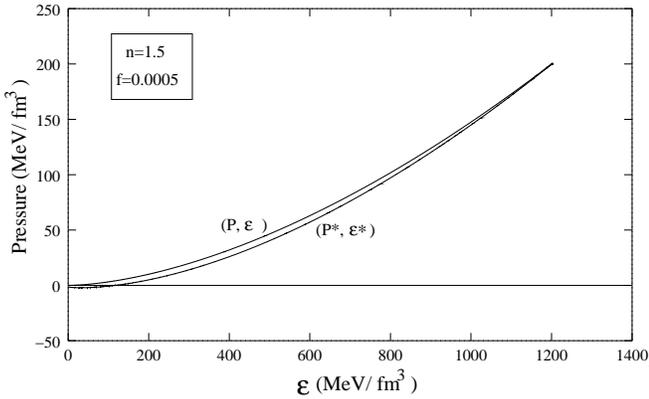,width=8.6cm}}\vskip .5cm
\caption{The pressure versus energy density plot for chargeless case and for
charge fraction $f=0.0005$.}
\label{fig:p-e-new5}
\end{figure}

\begin{figure}[htbp]\vskip .5cm
\centerline{\psfig{figure=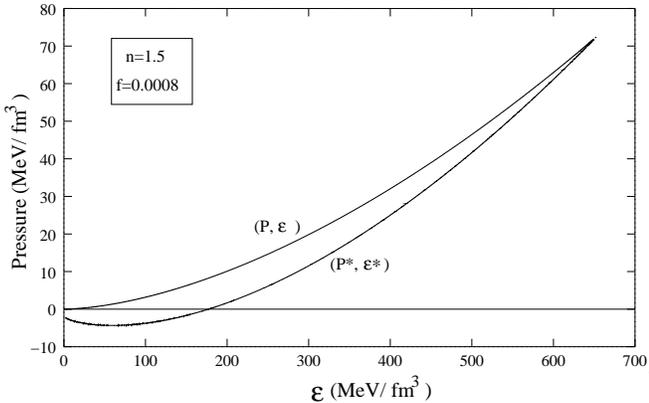,width=8.6cm}}\vskip .5cm
\caption{The pressure versus energy density plot for chargeless case and for
charge fraction $f=0.0008$.}
\label{fig:p-e-new8}
\end{figure}

>From  Eqs.  (\ref{eq:mnew})  and  (\ref{eq:dpdr}),  the effective
energy density and pressure of the system can be viewed as
\begin{eqnarray}
P^{*}&=&P-\frac{{\cal U}^2}{8\pi}~~~~~~~~~~~~~~~~~~\\
\nonumber
{\rm and}\\
\epsilon^{*}&=&\epsilon+\frac{{\cal U}^2}{8\pi}.
\end{eqnarray}
In  Figs. (\ref{fig:p-e-new5}) and  (\ref{fig:p-e-new8}) we  show two
different  charge fractions $f=0.0005$  and $f=0.0008$,  the effective
pressure  ($P^*$)  and energy  density  ($\epsilon^*$). The  effective
pressure drops  down to a negative  value, but even in  this case, the
first part of Eq. (\ref{eq:dpdr}) preserves its overall negative
value because ($M_{\rm tot}+4\pi r^3 P^*$) is positive. Thus the overall
 sign of the
pressure gradient $\frac{dP}{dr}$ is still negative as long as the
attractive gravitational term is larger than the  repulsive Coulombian one.
These figures show  that the effective EOS becomes  stiffer due to the
inclusion of charge and consequently allowing more mass in the star.

Also,  the  effective  pressure  directly  reduces the  value  of  the
negative  part of  $\frac{dP}{dr}$, but  the effective  energy density
increases the same  through the $ M_{\rm tot}$. This  goes on until the
effective  pressure becomes  so much  negative that  it  overcomes the
value of the $M_{\rm tot}$ and  this limits the formation of star with
higher charge fraction $f$.

\subsection{Balance between the Coulomb and gravitational forces}
In this section, we discuss the effects brought in the pressure gradient
from the matter energy and the Coulomb energy.
\begin{figure}[htbp]\vskip .5cm
\centerline{\psfig{figure=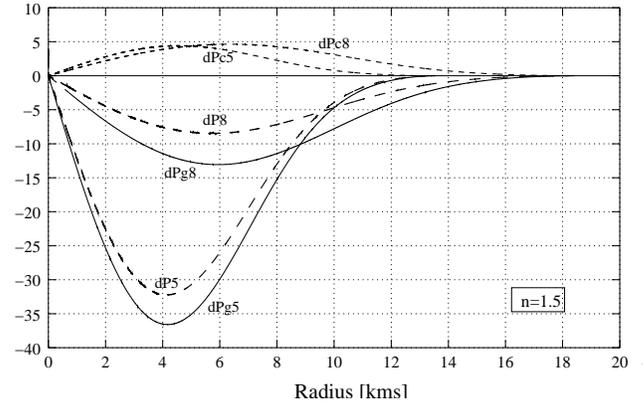,width=8.6cm}}\vskip .5cm
\caption{The Pressure gradient with the positive Coulomb part and the
gravitational negative part  is shown  here for two  different values of
the charge
factor $f$  (=0.0005 and 0.0008).  The total, $\frac{dP}{dr}$,  is also
plotted for  the same. The  annotations for the component  of pressure
for two  charges $f=0.0005$ and  0.0008 coming from  the $matter~part$
are dPg5 and dPg8 respectively, those from $Coulomb~part$ are dPc5 and
dPc8 respectively and the corresponding totals are dP5 and dP8.}
\label{fig:poly-dpdr}
\end{figure}
As mentioned  previously, the total  mass of the system  $M_{\rm tot}$
increases with  increasing charge because the  electric energy density
`adds on' to  the mass energy density. This change in  the mass is low
for  smaller charge  fraction  and going  up to  seven times  the value  of
chargeless case  for maximum allowed charge  fraction $f=0.0011$. This
effect however does not change the metric coefficient considerably when
compared to  the chargeless  case  (see Fig.  (\ref{fig:poly-l-r})). So,  the
ruling  term   in  Eq. (\ref{eq:dpdr})  is  the   factor  ($
M_{\rm tot}+4\pi  r^3(P-\frac{{\cal U}^2}{8\pi})$).  With the  increase of
charge,  the  value of  (P-$\frac{{\cal  U}^2}{8\pi}$) decreases,  and
hence   the  gravitational  negative   part  of   Eq.  (\ref{eq:dpdr})
decreases. The central  pressure is very high. So,  with the softening
of the pressure  gradient, the system allows more  radius for the star
until  it reaches  the surface  where the  pressure (and  the pressure
gradient too) goes to zero. We should stress that because $\frac{{\cal
U}^2}{8\pi}$ cannot be  too much larger than the  pressure in order to
maintain  $\frac{dP}{dr}$  negative as discussed before,  so   we  have
a  limit  on  the
charge. It is interesting that  this limit comes from the relativistic
effects of the gravitational force and not just only from the repulsive
Coulombian part.

This effect  is illustrated  in Fig. (\ref{fig:poly-dpdr})  where we
plot  both the positive  Coulomb part and the  negative matter
part  of the  pressure gradient. Note that, although we mention a
Coulomb part and a matter part, both  are coupled with each other
in the  sense that charge  changes the mass  and the mass  changes the
Coulomb part.  The plots are for  two values of  the charge fraction
$f=0.0005$  and  $f=0.0008$.  The  positive  part  of  $\frac{dP}{dr}$
maintains its almost constant value because the charge fraction $f$ is
the controller  of the same,  and in our  case, they differ by  a very
small percentage.  In the negative  part, the changes are  drastic and
are mainly brought by the effective pressure as we already discussed.

\section{Stability of charged stars and formation of
charged black holes}\label{sec:ns2bh}

\subsection{Stability considerations}

Here we will discuss the stability of these charged compact stars
taking into account the forces acting only on the charged
particles. For this, we need to compare our results with the stable
configuration of an almost neutral neutron star.

The basic argument to assume the charge neutrality of Newtonian stars,
or quasi-Newtonian stars (such as neutron stars), is based on the fact
that the total charge of the stars should lie below a certain limit
where the Coulomb repulsive force overwhelms the gravitational
attractive force at the surface of the star. This limiting value can
be viewed as the Coulomb force acting on a proton at the surface and
we have the limiting range as :
\begin{equation}
\frac{(Ze)e}{R^2} \le \frac{GMm}{R^2} \le \frac{G(Am)m}{R^2}
\label{eq:limit}
\end{equation}
where, $Ze$, $R$ and $M$ are the net charge of the star, its radius and mass
respectively and $m$  and $e$ are the mass and charge  of a proton. In
the above equation,  the mass of the star is  considered to be smaller
than  $Am$ because  of the  gravitational  binding of  the system.  In
gravitational units,
\begin{equation}
\frac{Z}{A} < \left(\frac{m}{e}\right)^2 < 10^{-36}.
\label{eq:lim-old}
\end{equation}

If we take into  account that there are approximately $A\simeq10^{57}$
baryons  in a neutron  star ($M\simeq M_\odot$), so  $ Z  < 10^{21},$  which
gives
\begin{equation}
Q \simeq Ze \simeq  100 ~ {\rm Coulomb}.
\label{eq:qold}
\end{equation}
This is the  limit on the net positive  charge already discussed in the
introduction. If  the star  had a  net negative  charge,  then the
electrons being the carrier of charge, the value of the limit would be
reduced by the factor $m_e/m$ \cite{glen00}.
Using the  solution for the TOV  equation in the
presence  of  electric fields,  which  we already solved in  the  previous
sections, we showed that  the balance of the forces, to make an element
of the fluid  at rest (hydrostatic equilibrium equation)  allows a
very large amount of charge.

In fact we showed that the maximum allowed charge is obtained when
$\rho_{\rm ch}\simeq\sqrt{G}\rho$ in natural units from the
assumption that more mass can hold more charge ($\rho_{\rm
ch}\propto\rho$).  Because of the large density found in
the compact stars like neutron stars ($\rho\simeq10^{15}\,$g/cm$^3$) we
can expect a large charge if $\rho_{\rm ch}$ is of the same order as
$\rho$.  Thus, $Q(r)\simeq \sqrt{G}M(r)$ and at the surface
of a neutron star with $M\simeq M_\odot$, the total charge is $
Q\simeq\sqrt{G}M_\odot\simeq10^{20}$ Coulomb and the intensity of the
electric field is of the order of $10^{21}-10^{22}$ V/m. In the case
of maximum charge, the relation $Q(r)=\sqrt{G}M(r)$ at the surface
shows that the ratio $Z/A\simeq 10^{-18}$ as compared to $10^{-36}$ in
Eq. (\ref{eq:lim-old}). This explains why we are having charges
$10^{18}$ times larger than that in relation (\ref{eq:qold}).  This
$Z$ is the net charge in the star and is the difference of the charged
particles with opposite sign ($Z\sim Z_{net}\sim |Z_+-Z_-|$). The
number of charges of equal and opposite sign is not at all limited.

However, not only a global balance of all the forces is enough to
guarantee the stability of these stars when we consider the forces
acting on each of the charged particles of that fluid. As the $Z/A$
ratio is $\simeq 10^{-18},$ the Coulomb force is $10^{18}$ times
larger than the gravitational one felt by a charged particle. This
will make the charged particles leave the system, making the star
unstable unless any other mechanism exists to bind together one
charged particle with $10^{18}$ neutral particles.

Our analysis reveals that the maximum amount of charge which can
be allowed in the system of charged compact stars is close to
unity (precisely 0.95) in geometrical units and can be compared
to the extremum case of charge in a Reissner-Nordstr\"om black hole, i.e.,
$Q/\sqrt{G}M=1.$ The stability analysis of the charged star from
the viewpoint of the force acting on a single charged particle
makes the star highly unstable, with the charge escaping from the
star within a very short time.  However, very little amount of
charge will escape before it collapses to a charged black hole.
This can be seen as follows : with the escape of little amount of
charge, the repulsive pressure from the Coulomb part diminishes,
but the attractive pressure from the matter part is not affected
very much. This is because the gravitational contribution of the
escaped charged particle (say proton) is practically zero as
compared to that from the 10$^{18}$ neutral particles. This leads
to a very critical scenario of disbalance of the global forces
inside the star, where the loss of a little amount of charge
makes the gravitational attractive pressure overwhelm the
residual repulsive Coulombian pressure and the whole system
collapses further to form black holes. But all the charge has not
yet escaped from the system and so, the residual charge gets
trapped in the black hole forming the charged black holes. This
picture is best suitable for the extreme charged case, but can be
true for smaller charged fractions also. We saw in Fig.
(\ref{fig:poly-e-m}), that the maximum stable mass for the
highest allowed charge fraction ($f=0.0011$) is as high as 9.7
$M_\odot$ and such high mass charged compact stars are more
favorable to collapse to Reissner-Nordstr\"om black holes.

\subsection{Discharge time}

Since the electric force on a charged particle
is much greater than the gravitational force
one can neglect the gravitational force, and
find that, from Newtonian theory, the
equation of
motion for a charged particle is
\begin{eqnarray}
\nonumber
m\, \frac{\partial^2 r}{\partial t^2}=\frac{Q\,q}{r^2}
\end{eqnarray}
where $m$ and $q$ are the mass and charge of the particle,
and $Q$ is the net charge of the star.
Thus, the lifetime to discharge the star $t_{\rm discharge}$
is given in Newtonian theory by
$ t_{\rm discharge\, Newt} \simeq 
\frac{1}{c}\sqrt{\frac{R^3}{G M/c^2}\frac{m}{q}},$
where we have used the relation $Q\simeq\sqrt{G}M$. For $R\sim10\,$km,
$GM/c^2 ~\sim 3\,$km and $m/q \sim 10^{-18}$, we have, $t \sim
10^{-13}\,$s. However this time scale is much shorter than the
time a particle takes to traverse the radius of the star. This
means that the charged particle acquires quickly a velocity near
the speed of light and it will travel throughout the star with
this speed. Thus, the discharging time is of the order of the
crossing time, i.e.,
\begin{eqnarray}
\nonumber
t_{\rm discharge} \simeq \frac{R}{c}
\end{eqnarray}
For a star with $R\sim10\,$km
$t_{\rm discharge} \simeq 10^{-5}\,$s.

We have not taken into account that the particle collides with
other particles with a mean free path of the order of one fermi.
However, the trajectory of the charged particle is not going to
be a random walk trajectory, since the electric force is radial.
Thus, this collision process might increase a bit the discharging
time, but not a lot. A detailed evaluation is however a
very complicated process and is beyond the scope of the present
paper.

\section{Conclusions}\label{sec:conc}
Our analysis shows that the amount of charge contained in a dense
system like a compact star can be very high and several orders of
magnitude larger than those calculated by classical balance of
forces at the surface of a star. This amount of charge mainly
comes from the high density of the system since $\rho_{\rm
ch}\propto\rho$. We showed that the charge can be as high as
10$^{20}$ Coulomb to bring in any change in the mass-radius
relation of the star, yet remaining stable as long as one
considers only the hydrostatic equilibrium and the global balance
of forces and not considering the individual forces on the
charged particles. In our study, we used a polytropic EOS for our
compact star with a choice of parameters such that the system is
close to the realistic neutron stars. We showed that in the
critical limit of the charge contained in the system, the maximum
mass stable star forms in a lower density regime, however
compactness keeps on increasing.  We have also studied the change
in the pressure gradient due to the effect of the charge
contained in the system. It was expected from the classical
picture of forces involved, that the repulsive force of the
charged particles will add up to the internal pressure of the
system and the entire repulsive force will be balanced by the
gravitational force of the system.  However, we can see from
Fig.  (\ref{fig:poly-dpdr}) that the contributions from the
charged particles are helping to soften the gravity part of the
pressure gradient, thus allowing more matter to stay in the
system. The second term in the right hand side of Eq.
(\ref{eq:dpdr}) is positive always and does not depend on the
nature of the charge whether it is positive or negative. The
pressure gradient must be a negative term and hence
$\frac{dP}{dr}$ is softened by the effect of the presence of
charge. The net effect is the gravitational force which tries to
collapse the system is held further away by the Coulomb force,
which, in the absence of the gravity would have exploded the
system.

This imbalance of the pressure actually will come to play when some of
the charge will leave the star due to the Coulombian repulsion acting
on a single charged particle. As we discussed in section
\ref{sec:ns2bh}, the attractive gravitational pressure will then
become more than the repulsive total kinetic and Coulombian pressure,
as a result of which the star may collapse to a charged black hole.

We have studied the electric field inside the maximum mass star
allowed by a certain charge fraction. We found that the field attains
a maximum value for certain amount of charge fraction and then
decreases for higher charge fractions (Fig.  (\ref{fig:poly-u-r})).
This is also interesting because normally it was expected that the
increase in the charge fraction would increase the field also. But as
our charge distribution varies directly with the matter density, so
the formation of the maximum mass star in the lower density regime for
high charge fractions, reduces the electric field. The mass of the
stars (Eq. (\ref{eq:mnew})) in these critical field limit is very high
as compared to the chargeless case and have a contribution from the
electrostatic energy density, but this is small compared to the
contribution of the matter density.

As pointed by Bekenstein \cite{bek71}, with this amount of huge charge
(as we have in our system), the electric field produced will be too
high and give rise to pair production.  This effect results in self-diminishing the
electric field and thus the system will destabilize \cite{sp01}.  However, we can
say that the critical field limit calculation has been done
in vacuum, and it is not at all
clear how the field will behave in a dense system. Recent
observations have revealed that there are magnetars, stars with
strong rotating fields, which have a
magnetic field as high as $10^{15}$ Gauss. The critical
magnetic field limit for pair creation in vacuum is $10^{13}$
Gauss. So, if such high fields really exist in a highly dense magnetar
and these stars are stable, then the critical field limit needs to be
modified for high density matter. Putting aside this debate of the
critical field limit, we have checked only the behaviour of the system
in the presence of a very high charge. We have found that the global
balance of forces between the matter part and the electrostatic part can support
a huge amount of charge. In addition, we have argued
that the solution is unstable,  and a decrease in
the electric field can create an enormous
imbalance, resulting in the collapse of the charged star to a charged
black hole. Finally, these charged stars should be very short lived,
they would exist within the very short period between the supernova
explosion and the formation of the charged black holes.

\acknowledgements
{SR likes to thank the FAPERJ, ALE to CAPES for research support and
MM for the partial CNPQ support. We also thank P. Alberto,
 M. Fiolhais, H. J. M. Cuesta
and K. Dechoum for some useful discussions. JPSL and VTZ thank
the hospitality and support of Observat\'orio
Nacional do Rio de Janeiro. We also thank very much to the anonymous referee for some useful
comments and suggestions which immensely improved the paper.}

\bigskip\bigskip
\appendix*
\section{Conversion to real units}
It is important  to mention here the basic units used in our approach
and their conversions from these units to the real units of the charge
and  fields like  Coulomb and  Volt/meter respectively.  We  used the
charge density ($\rho_{\rm ch}$) proportional to the energy density of the
system ($\rho$), which in turn is in  MeV/fm$^3$, with a factor $f$ which has
dimensional units. From the fine structure constant
$\alpha = \frac{e^2}{\hbar c} =\frac{1}{137}$,
we get a relation $\rm 1 ~statCoulomb (1~ e.s.u.)=2.51\times10^9
~[MeV~fm]^{1/2}$. Using the
conversion factor $\rm 1 ~Coulomb=2.998\times10^9 ~e.s.u.$
for the charge we arrive at
\begin{equation}
1 \,{\rm Coulomb} \simeq 0.75 \times 10^{19}\, {\rm [MeV\, fm]}^{1/2}.
\end{equation}
Also, the units of $\frac{{\cal U}^2}{8\pi}$ have to be the same as
the units of pressure. Initially, we have that the units of pressure are
[MeV/fm$^3$]. This indicates that ${\cal U}$ will have
$\rm [MeV/fm^3]^{1/2}$ units. Working out the relations, we find
\begin{equation}
1 {\rm \left[\frac{MeV}{fm^3}\right]}^{1/2}=1.2 \times 10^{21}
\, {\rm V/m}.
\end{equation}
Also,
\begin{eqnarray}
\nonumber
1 {\rm \left[\frac{MeV}{c^2 fm^3}\right]}=1.78\times 10^{12}
\,{\rm g/cm^3}.
\end{eqnarray}

\end{document}